\documentclass[american,12pt]{article}
\usepackage[T1]{fontenc}
\usepackage{amssymb}
\usepackage{graphicx}
\usepackage{pdfpages}

\usepackage{scicite}
\usepackage{times}

\newenvironment{sciabstract}{%
\begin{quote} \bf}
{\end{quote}}

\newcounter{lastnote}

\title{Ballistic miniband conduction in a graphene superlattice}
\author
{Menyoung Lee,$^{1}$ John R.\@ Wallbank,$^{2}$ Patrick Gallagher,$^{1}$ Kenji Watanabe,$^{3}$\\
Takashi Taniguchi,$^{3}$ Vladimir I.\@ Fal'ko,$^{2,4}$ and David Goldhaber-Gordon$^{1\ast}$\\
\normalsize{$^{1}$Department of Physics and Geballe Laboratory for Advanced Materials,}\\
\normalsize{Stanford University, Stanford, California 94305, USA}\\
\normalsize{$^{2}$National Graphene Institute, University of Manchester, Booth St. E, Manchester,}\\
\normalsize{M13 9PL, UK}\\
\normalsize{$^{3}$National Institute for Materials Science, 1-1 Namiki, Tsukuba 305-0044, Japan}\\
\normalsize{$^{4}$School of Physics \& Astronomy, University of Manchester, Oxford Road, Manchester,}\\
\normalsize{M13 9PL, UK}\\
\normalsize{$^\ast$To whom correspondence should be addressed;
E-mail:  goldhaber-gordon@stanford.edu.}
}
\date{}

\begin{document}
\baselineskip24pt
\maketitle

\begin{sciabstract}
Rational design of artificial lattices yields
effects unavailable in simple solids \cite{Esaki1970}, and vertical
superlattices of multilayer semiconductors are already used in optical
sensors and emitters \cite{Esaki1974,Waschke1993,Faist1994}. Manufacturing
lateral superlattices remains a much bigger challenge \cite{Weiss1989a,Carmona1995,Albrecht1999a},
with new opportunities offered by the use of moir\'{e} patterns in van
der Waals heterostructures of graphene and hexagonal crystals such
as boron nitride (h-BN) \cite{Xue2011a,Decker2011,Yankowitz2012}.
Experiments to date have elucidated the novel electronic structure
of highly aligned graphene/h-BN heterostructures \cite{Yankowitz2012,Ponomarenko2013a,Dean2013a,Hunt2013a,Yu2014b,Shi2014d},
where miniband edges and saddle points in the electronic dispersion
can be reached by electrostatic gating. Here we investigate the dynamics
of electrons in moir\'{e} minibands by transverse electron focusing, a
measurement of ballistic transport between adjacent local contacts
in a magnetic field \cite{Tsoi1974}. At low temperatures, we observe
caustics of skipping orbits extending over hundreds of superlattice
periods, reversals of the cyclotron revolution for successive minibands,
and breakdown of cyclotron motion near van Hove singularities. At
high temperatures, we study the suppression of electron focusing by
inelastic scattering.
\end{sciabstract}

In solids, the quantum nature of electrons generates band structure
which controls conduction and optical properties. Similarly, longer-period
superlattices in solids possess minibands that disperse at a finer
energy scale over a reduced Brillouin zone, enabling phenomena such
as negative differential conductance and Bloch oscillations \cite{Esaki1970,Esaki1974,Waschke1993}.
Two-dimensional (2D) electron systems could be a promising platform
on which to tailor superlattice minibands. Yet fabricating long-range
periodic patterns that strongly modulate the potential to form well-separated
minibands without undermining the material quality and electron coherence
remains challenging. Most experiments on laterally patterned
semiconductor heterostructures have revealed classical commensurability
effects \cite{Weiss1989a,Carmona1995,Sandner2015} which do not require
phase coherence, and only subtle features have been attributed to
miniband formation \cite{Albrecht1999a}.

The arrival of high-quality graphene/h-BN van der Waals heterostructures
with misalignment angle below $1^{\circ}$ \cite{Xue2011a,Decker2011}
has drastically changed the situation. In such systems, the periodic
potential for electrons in graphene is imposed by the hexagonal moir\'{e}
pattern generated \cite{Ortix2012,Kindermann2012a,Wallbank2013} by
the incommensurability and misalignment between the two crystals.
Formation of minibands for Dirac electrons has been demonstrated by
magnetotransport \cite{Ponomarenko2013a,Dean2013a,Hunt2013a}, as
well as scanning tunneling \cite{Yankowitz2012}, capacitance \cite{Yu2014b},
and optical \cite{Shi2014d} spectroscopies. The connection between
the miniband dispersion $\varepsilon(\vec{k})$ and transport properties
is established by the equations of motion for an electron in an out-of-plane
magnetic field $\vec{B}=B\hat{z}$, 
\begin{equation}
\hbar\vec{v}=\frac{\partial\varepsilon}{\partial\vec{k}},\ \hbar\dot{\vec{k}}=-e\vec{E}+eB\hat{z}\!\times\!\vec{v},\label{eq:semi}
\end{equation}
where the relation between carrier velocity $\vec{v}$ and momentum
$\hbar\vec{k}$ is approximately
$\vec{v}=v\vec{k}/k$ ($v\approx10^{6}\mbox{\thinspace m/s}$), close
to the Dirac point of graphene's spectrum \cite{Yu2014b,Ortix2012,Kindermann2012a}.

The shape of the cyclotron orbit in a 2D metal is a $90^{\circ}$
rotation of the shape of the Fermi surface, and the carrier revolves
along it clockwise or counterclockwise. Electron trajectories
near the boundary of a metal open into skipping orbits \cite{Bohr}
which drift in the direction determined by the effective charge of
the carrier. These skipping orbits bunch along caustics \cite{Beenakker1988a,Davies2012,Patel2012},
leading to the transverse electron focusing (TEF) effect \cite{Tsoi1974}.
Experimentally, TEF takes place when the magnetic field is tuned such that
caustics of skipping orbits, emanating from an emitter $E$, end up
at a collector $C$, located at position $x=L$ along the boundary. Then
a voltage $V_{C}$ is induced at $C$, proportional to the current $I_{E}$
injected into $E$. Fig.\@ 1B illustrates skipping orbits and caustics
in a material with an isotropic Fermi surface, such as unperturbed
graphene near the Dirac point, where TEF occurs for $B=B_{j}\equiv\frac{2j\hbar k_{F}}{\pm eL}\mbox{ (for \ensuremath{j\!=\!1,2,\!...})}$.
An equidistant series of peaks (oscillations) appears in the focusing
``spectrum''---the non-local magnetoresistance $V_{C}/I_{E}(B)$
(Fig.\@ 1C), from which the Fermi momentum $\hbar k_{F}$ and the
sign of effective charge $\pm e$ may be inferred. TEF was initially
used to study the Fermi surfaces of bulk metals \cite{Tsoi1974,VanSon1987a},
and was later extended to 2D systems \cite{Beenakker1988a}, including
graphene \cite{Taychatanapat2013b}.

\begin{figure}[!ht]
\centering\includegraphics[width=0.8\textwidth]{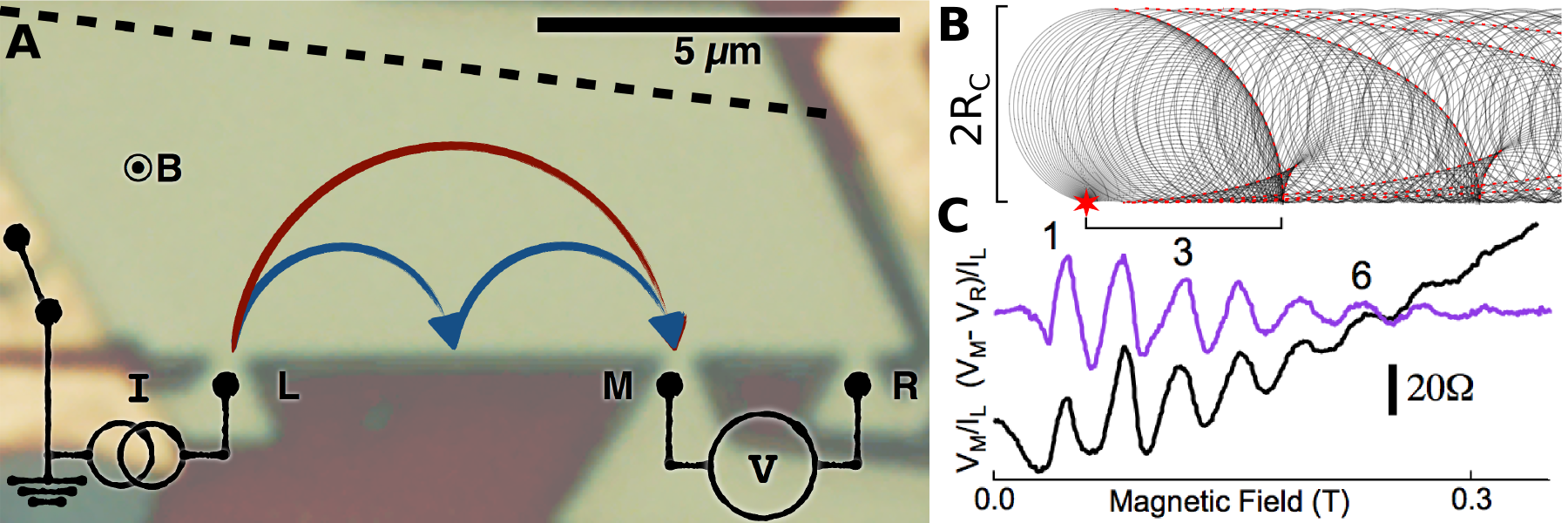}
\begin{minipage}{0.95\columnwidth}
\smallskip \small \baselineskip20pt
\textbf{Fig.\@ 1.} (\textbf{A}) Schematic of the experiment overlaid on a photo of the
device. The h-BN/graphene/h-BN/bilayer graphene heterostructure is
green, the SiO$_{2}$ substrate is purple, and the dashed line denotes
the upper boundary of the graphene flake. Electrical measurement configuration
applied to obtain data in Fig.\@ 2B: the two leftmost contacts are
grounded to act as absorbers. We inject current into the left local
contact $L$ and measure the voltage difference between two local
contacts, $M$ and $R$. Arrows depict skipping orbits a hole would
take if injected at normal incidence with $B=B_{1}\equiv\frac{2\hbar k_{F}}{eL}$
(red) or $B_{2}\equiv\frac{4\hbar k_{F}}{eL}$ (blue). (\textbf{B})
Simulated ensemble of skipping orbits emanating from an emitter (red
star). Electron trajectories bunch along caustics (red dashed curves)
and focus onto an equidistant array of points at the boundary. Scale
markers show the cyclotron diameter $2R_{C}=\frac{2\hbar k_{F}}{eB}$.
(\textbf{C}) Transverse electron focusing (TEF) spectra collected
at a single voltage probe $M$ ($V_{M}/I_{L}(B)$, lower trace), and
differentially between voltage probes $M$ and $R$ ($\left(V_{M}\!-\!V_{R}\right)\!/I_{L}(B)$,
upper trace), with $n\!=\!-1.1\!\times\!10^{12}\mbox{\thinspace cm}^{-2}$.
The first, third, and sixth focusing peaks are labeled. Taking the
differential measurement of the spectrum does not shift peak positions,
because the device geometry partially shields $R$ from being reached
by skipping orbits from $L$, such that TEF oscillations of $V_{R}$
are much weaker.
\end{minipage}
\end{figure}

Here we report the observation of TEF in a moir\'{e} superlattice at the
interface between graphene and h-BN in a van der Waals heterostructure
(from top to bottom) h-BN/graphene/h-BN/bilayer graphene assembled on
an SiO$_{2}$ substrate. One of the h-BN layers (we do not know which)
is aligned with graphene to better than $1^{\circ}$, forming a moir\'{e}
pattern with a 14\,nm period. We use the bilayer graphene as an electrostatic
gate, tuning electron density in the superlattice by applying voltage
$V_{g}$ to it. The device, depicted in Fig.\@ 1A, has three
etched local contacts along the linear sample boundary. Two other
ohmic contacts are grounded and act as absorbers. We measure the
multi-terminal, non-local resistance $\left(V_{M}\!-\!V_{R}\right)\!/I_{L}$
at our base temperature $T=T_{base}=1.55$\,K. Figure 2B is the resulting
map of $\left(V_{M}\!-\!V_{R}\right)\!/I_{L}$ as a function of $B$
and $V_{g}$, exhibiting electron focusing spectra and their evolution
as a function of electron density. When the Fermi level in graphene
is close to the Dirac point at $V_{g}=-0.4$\,V, the superlattice
spectrum is almost isotropic, and $k_{F}=\sqrt{\pi |n|}$. Hence the
focusing spectra show TEF oscillations with peaks at $B_{j}=\frac{2j\hbar}{\pm eL}\sqrt{\pi |n|}$
(dashed curves in Fig.\@ 2B) as in unperturbed graphene \cite{Taychatanapat2013b}.
The observation of TEF confirms that electrons travel
ballistically from emitter to collector.

\begin{figure}[!ht]
\centering\includegraphics[width=1\textwidth]{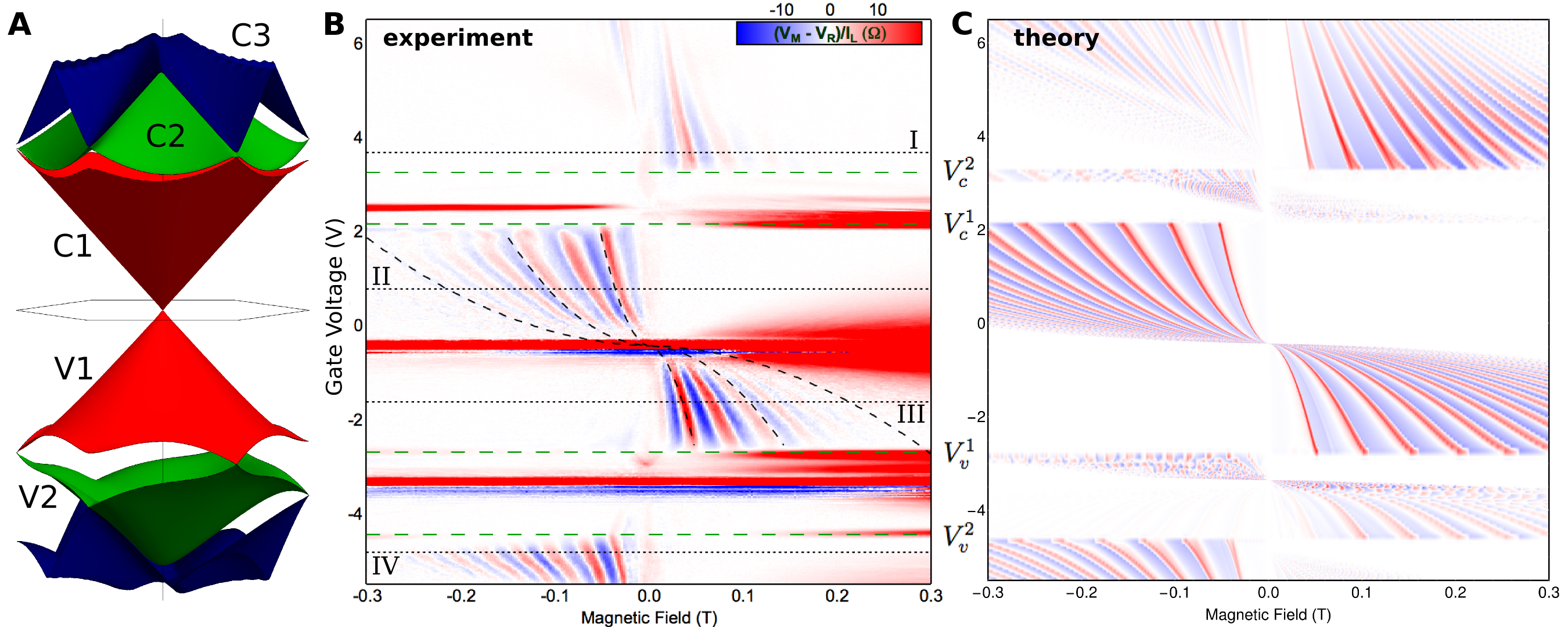}
\begin{minipage}{0.95\columnwidth}
\smallskip \small \baselineskip20pt
\textbf{Fig.\@ 2.} (\textbf{A})
Calculated miniband structure of the graphene/h-BN superlattice.
Each miniband in which we observe TEF is given a label as shown. This
dispersion results from a symmetric moir\'{e} perturbation: $\epsilon^{+}=17$\,meV
and $\epsilon^{-}=0$\,\,meV. This choice gives the best match
to experimental data, out of a two-parameter family (Supplementary Materials).
(\textbf{B}) TEF spectra as a function of gate voltage $V_{g}$. The
plotted ratio $\left(V_{M}\!-\!V_{R}\right)/I_{L}$ is measured as
depicted in Fig.\@ 1A. Dashed curves: $B_{1}$, $B_{3}$, and $B_{6}$, which are
some of the peak positions expected when the system is near the Dirac point.
Dashed lines indicate the abrupt termination of TEF due to the breakdown
of cyclotron motion at saddle point van Hove singularities. They are
labeled by the miniband in which the breakdown occurs, e.g.\@ $V_{c}^{1}$
for the breakdown of cyclotron motion in C1. Dotted lines:
selected densities, I, II, III, and IV, which place the Fermi level
in minibands C2, C1, V1, and V2, respectively. (\textbf{C}) TEF spectra
as a function of $V_{g}$, calculated from the dispersion in (A) and Eq.\@ \ref{eq:semi}.
\end{minipage}
\end{figure}

At higher densities of about four electrons (or holes) per moir\'{e} unit
cell, the Fermi level is near the first minibands' outer edges,
and TEF spectra reflect the modification of electronic
states by the superlattice potential. A candidate miniband structure
from the model family proposed in \cite{Wallbank2013} is rendered
in Fig.\@ 2A, where we label relevant minibands. In addition to TEF
of electrons in C1 and holes in V1, we detect focusing of holes in
C2 and electrons in V2 and C3. Carrier dynamics in the form of skipping
orbits and caustics are represented using ensembles of simulated electron
trajectories in Fig.\@ 3. The map of measured TEF spectra, Fig.\@ 2B,
matches very well with the theoretically simulated spectra in Fig.\@ 2C,
obtained by applying Eq.\@ \ref{eq:semi} to the electrons emitted
into the minibands of Fig.\@ 2A from a local emitter at the sample
edge (the calculation is fully described in Supplementary Materials).
TEF oscillations abruptly terminate at gate voltage values
$V_{v}^{1},V_{c}^{1},V_{v}^{2},\mbox{and }V_{c}^{2}$,
which coincide with the passing of the Fermi level across
the saddle point van Hove singularities at which the constant energy
contour of the miniband dispersion percolates across all repeated
Brillouin minizones. At these saddle points, cyclotron orbits experience
an extreme variant of magnetic breakdown termed orbital switching
\cite{Markiewicz1994}---opening up into run-away trajectories such
that electrons do not drift along the edge of the sample following
skipping orbits. In the ranges $V_{v}^{2}\!<\!V_{g}\!<\!V_{v}^{1}$
and $V_{c}^{1}\!<\!V_{g}\!<\!V_{c}^{2}$, the Fermi surface consists of small
and highly anisotropic pockets just above or below the secondary
Dirac points, thus even the theoretically calculated pattern in Fig.\@ 2C
is weak and complex. Accordingly, we don't observe prominent TEF oscillations
over these ranges as we do for the biggest pocket of each miniband.
For $V_{g}\!>\!V_{c}^{3}$,
where $V_{c}^{3}$ is the lower band edge of C3, the electron-like pocket of C3 overlaps
in energy with hole-like pocket of C2, leading to TEF oscillations for both
signs of $B$.

\begin{figure}[!ht]
\centering\includegraphics[width=0.7\textwidth]{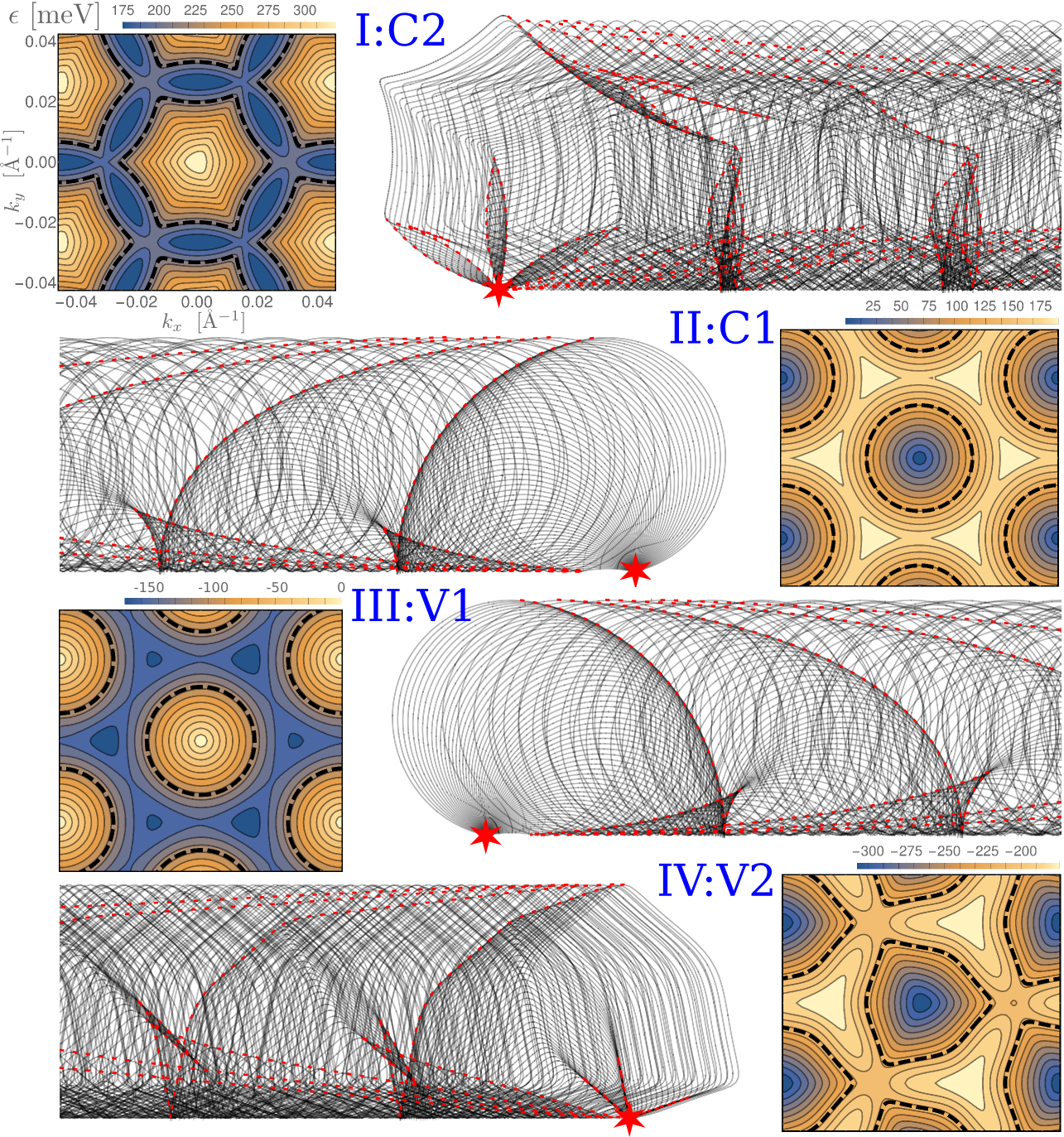}
\begin{minipage}{0.95\columnwidth}
\smallskip \small \baselineskip20pt
\textbf{Fig.\@ 3.} Representative ensembles of simulated skipping orbits emanating from
an emitter (red star) at the boundary of the graphene/h-BN superlattice
possessing the miniband dispersion of Fig.\@ 2A, for selected electron
densities I, II, III, and IV marked in Fig.\@ 2B. The corresponding
Fermi surfaces are in minibands C2, C1, V1, and V2, respectively,
and each one is drawn as a thick, dashed constant-energy contour on
the color map of the dispersion. The magnetic field
points out of the page, so electron-like carriers turn counter-clockwise
and their skipping orbits drift left, and hole-like carriers do
the opposite. Red dashed curves mark caustics.
\end{minipage}
\end{figure}

The saddle points $V_{v}^{1},V_{c}^{1},V_{v}^{2},\mbox{and }V_{c}^{2}$
can be directly compared to miniband models. We tested the observed
ratios $\frac{V_{v}^{1}-V_{v}^{2}}{V_{c}^{1}-V_{v}^{1}}$ and $\frac{V_{c}^{2}-V_{c}^{1}}{V_{c}^{1}-V_{v}^{1}}$
against predictions for a family of moir\'{e} perturbations in graphene/h-BN
heterostructures parameterized by $\epsilon^{+}$ and $\epsilon^{-}$,
the respective strengths of spatially symmetric and antisymmetric
interlayer couplings from graphene to the boron and nitrogen sites
of h-BN (complete definition in Supplementary Materials) \cite{Chen2014a}.
The best match to experimental data results from taking a symmetric
moir\'{e} perturbation $\epsilon^{+}\approx17\,\mbox{meV},\epsilon^{-}\approx0$
(Supplementary Materials). This parameter choice is used to calculate
the miniband structure, electron dynamics, and TEF maps shown in
Figs.\@ 2 and 3. Our value for $\epsilon^{+}$ is similar to previous estimates
by optical spectroscopy \cite{Shi2014d}.
We furthermore set an upper limit, $\left|\epsilon^{-}\right|<3$\,meV, on the antisymmetric
potential, whose absence had previously simply been assumed \cite{Shi2014d}.

\begin{figure}[!ht]
\centering\includegraphics[width=0.8\textwidth]{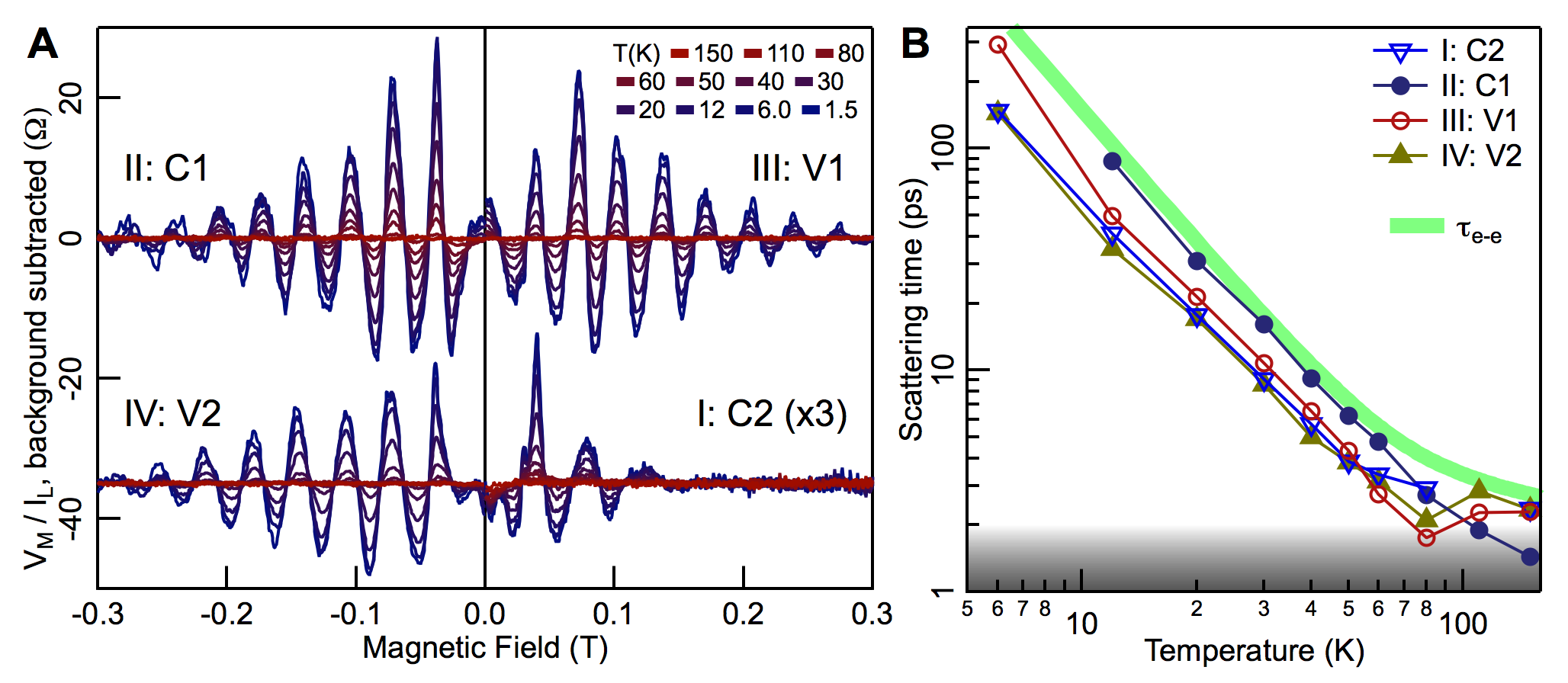}
\begin{minipage}{0.95\columnwidth}
\smallskip \small \baselineskip20pt
\textbf{Fig.\@ 4.} (\textbf{A}) Temperature dependence of TEF spectra,
$V_{M}/I_{L}\left(B\right)$ minus a smooth background, for the electron densities
I, II, III, and IV marked in Fig.\@ 2B. (\textbf{B}) Symbols: effective scattering times
$\tau(T)$ extracted from the suppression of TEF oscillations upon increasing the
temperature, for the same four densities as (A). Curve: theoretical
scattering time $\tau_{e-e}$ related to the electron-electron interaction.
The detection limit set by noise is shaded.
\end{minipage}
\end{figure}

We can learn more about carrier dynamics, in particular the effect
of their scattering, by examining the temperature dependence of TEF
oscillations \cite{VanSon1987a}. Throughout the probed temperatures and densities,
the suppression of TEF upon heating (see Fig.\@ 4A) is faster than what could be
expected from merely thermal broadening of injected electron momenta, as 
$|k-k_{F}|\!\sim\!\frac{k_{B}T}{\hbar v_F}\ll k_{F}$. 
For quantitative analysis, we determine the area $A_{1}$ under the first ($j\!=\!1$)
focusing peak and interpret the ratio $A_{1}\!\left(T\right)/A_{1}\!\left(T_{base}\right)$
as the fraction of electrons $\sim\!e^{-\pi L/2v_F \tau}$ that propagated ballistically from the
emitter to the collector, along the semicircle of a cyclotron trajectory of length
$\frac{\pi L}{2}$ that touches the caustic near the collector,
despite the electrons scattering with a characteristic time constant $\tau$. 
In Fig.\@ 4B, we show the temperature dependence of this effective scattering time,
extracted from the data using the formula 
$\tau\left(T\right)\!=\!-\frac{\pi L}{2v_F}/\log\frac{A_{1}\left(T\right)}{A_{1}\left(T_{base}\right)}$. 
The experimentally observed dependence $\tau(T) \propto T^{-2}$ points toward an
electron-electron ($e$-$e$) scattering mechanism for the suppression of
TEF oscillations upon heating, the same mechanism
associated with the evolution of electronic transport 
from ballistic to the viscous regime \cite{Torre2015a,Principi2015a,Bandurin2015}.
Theoretical analysis of spreading of a narrow beam of electrons due to the low-angle
electron-electron scattering processes, performed in Supplementary Materials 
using Thomas-Fermi-screened $e$-$e$ interaction, shows 
that for $T \lesssim T_*$ (where $k_B T_* = 2v_F \sqrt{\frac{k_F}{\pi L}}$),
the decay of TEF signal can be described by
\begin{equation}\label{eq:tau_approx}
 \tau^{-1}_{e-e} \approx \frac{(k_BT)^2}{2\hbar v_F k_F} \log\left(  \frac{3.6 L}{w}  \right) ,
\end{equation}
where $w$ is the width of the emitting and collecting contacts.
The theoretically calculated values of scattering times are shown in Fig.\@ 4B,
including the theoretically predicted crossover to a slower scattering rate for
$T > T_*$ (Supplementary Materials). 
As these calculations \emph{with no free parameters} match the experimentally
found values, it is tempting to conclude that $e$-$e$ scattering is the dominant
mechanism for suppression of TEF. Electron-phonon scattering, however, may also
play a role \cite{Hwang2008c}. Characterization of the phonon spectrum and
electron-phonon coupling is required to quantify that effect \cite{Slotman2014};
the key parameters have been experimentally determined for graphene on
SiO$_{2}$/Si but not yet for graphene/h-BN heterostructures.

The direct observation and manipulation of ballistic transport is
a powerful probe of the low-energy physics of an electron system;
unlike in optical spectroscopies, the quasiparticles freely propagate
through the time of flight ($\sim\!10$\,ps here). Our experiment
elucidates the key basic features of miniband electron dynamics in
a moir\'{e} superlattice, and points toward fertile ground for further
explorations of novel transport effects. For instance, the saddle
point van Hove singularities could host exotic effects caused by enhanced
electron-electron interactions \cite{Nandkishore2012}, and valley-contrasting
physics could be accessed by taking advantage of the severe trigonal
warping of minibands \cite{Garcia-Pomar2008a}. For technology, such
a clear validation of the miniband properties suggests that graphene/h-BN
and perhaps other moir\'{e} superlattices may form a practical platform
for new devices based on miniband physics.

\paragraph*{Acknowledgments}
We thank A.\@ L.\@ Sharpe and G.\@ Pan for technical assistance and
W.\@ A.\@ Goddard, M.\@ S.\@  Jang, H.\@  Kim, A.\@  Maharaj, S.\@  Goswami,
E.\@ J.\@ Heller, L.\@ S.\@ Levitov, J.\@ C.\@ W.\@ Song
and A.\@ Benyamini for discussions.
Work done at Stanford was funded in part by
the Air Force Office of Scientific Research, Award No.\@ FA9550-16-1-0126,
and by the Gordon and Betty Moore Foundation through Grant No.\@ GBMF3429,
and performed in part in the Stanford Nano Shared Facilities (SNSF).
ML was supported by Stanford Graduate Fellowship and Samsung Scholarship.
JRW and VIF were supported by ERC Synergy Grant Hetero2D and
the EU Graphene Flagship Project.



\section*{Supplementary material}

# 1 Materials and Methods

All measurements were performed in the helium vapor space of a flow cryostat with a superconducting magnet. TEF spectra were measured by a lock-in amplifier, sourcing a small 263 Hz, 50 nA rms alternating current into the emitter contact, which ensured that the voltage drop at the contact was always less than $k_B T/e$.

## 1.1 Sample Fabrication

The two-dimensional system we investigated was a heterostructure of monolayer graphene encapsulated in hexagonal boron nitride (h-BN), with a bilayer graphene back gate. Flakes of monolayer and bilayer graphene (from highly oriented pyrolytic graphite, Momentive Performance Materials ZYA grade) and of h-BN (from single crystals synthesized at high pressure and high temperature) were exfoliated mechanically using tape (3M Scotch), deposited on oxidized (90 nm oxide) silicon (WRS Materials) substrates, and identified by inspection of optical microscope images. The flakes of h-BN had thicknesses $\approx 30$ nm. The heterostructure was assembled by a top-down pick-up technique using a temperature-sensitive adhesive film of polypropylene carbonate (Sigma-Aldrich) mounted on a clear backing of Polydimethylsiloxane (Dow Corning Sylgard 184), as described in (*36*). The assembled heterostructure was deposited on a chip of oxidized (300 nm oxide), degenerately doped silicon. We did not intentionally orient the flakes or anneal the final heterostructure to align the crystal axes as other authors have done (*37, 38*). The hetrostructure had no h-BN step edges or segregated bubbles over or under the device area. All patterns were defined by electron-beam lithography, using beams of 10 or 30 keV energy to expose the resist PMMA 950 A4 or A5 (MicroChem). The exposed pattern was developed by a chilled 1:3 mixture of water and isopropanol, in order to avoid swelling the resist which can lead to cracking or delamination from the h-BN surface (*39*). The device geometry, which includes three narrow local contacts along a linear boundary for carrier reflections, was defined

by direct reactive ion etching in a 150 mTorr plasma of $\text{CHF}_3/\text{O}_2$ with flow rates 50 and 5 sccm, respectively. The local contacts were etched to a nominal width of 250 nm, and adjacent pairs were separated by lengths 2.5 and 6.3 $\mu$m. Ohmic contacts to the device as well as contact to the bilayer back gate were formed by electron-beam evaporation of Cr/Au electrodes onto the edge of graphene exposed by the etch, as in (*36*).

## 1.2  Miniband theory and semiclassical model

The miniband structure was calculated for a fully aligned ($\theta = 0$) graphene/h-BN heterostructure by numerically diagonalizing the Hamiltonian (*20, 28*),

$$\hat{H} = v\vec{p}\cdot\vec{\sigma} + \frac{1}{2}\left(\epsilon^+ f_+ - \epsilon^- f_-\right) - \frac{\xi\sqrt{3}}{2}\sigma_3\left(\epsilon^+ f_- + \epsilon^- f_+\right) - \frac{\xi\vec{\sigma}}{b}\cdot\left[\hat{z}\times\nabla\left(\epsilon^+ f_- + \epsilon^- f_+\right)\right]. \quad \text{(S.1)}$$

Here $\sigma_i$ are Pauli matrices, acting on Bloch states $(\phi_{AK}, \phi_{BK})^T$ in the $K$ valley ($\xi = 1$) and $(\phi_{BK'}, -\phi_{AK'})^T$ in the $K'$ valley ($\xi = -1$), and $v$ is the Dirac velocity. Functions $f_\pm = \sum_m (\pm 1)^{m+\frac{1}{2}} e^{i\vec{b}_m\cdot\vec{r}}$ are written using the six shortest Bragg vectors of the moiré superlattice, $\vec{b}_{m=0,\cdots,5} = R_{m\pi/3}(0, b)$, where $R_\varphi$ describes anticlockwise rotation by angle $\varphi$, and we use $b = 0.053\,\text{Å}^{-1}$ for the fully aligned heterostructure. The band structure parameters $\epsilon^+$ and $\epsilon^-$ represent the strengths of the inversion symmetric and antisymmetric components of the moiré superlattice perturbation; see Eqs. S.2 and S.3 for the definition. Fig. S1 and corresponding text below shows that fitting this model to experimental data yields estimates $\epsilon^+ = 17\,\text{meV}$ and $\epsilon^- = 0$.

To numerically calculate the TEF spectra in Fig. 2C, we model the device shown in Fig. 1A as a wide graphene/h-BN strip with an emitter and two collectors placed along the lower boundary. We choose to orient the boundaries along the $x$-axis set by the direct lattice vectors of moiré pattern, but find that the main features in the TEF spectra are insensitive to this choice (Fig. S2 below and corresponding text). The emitter is modeled as a small patch of graphene/h-BN with a slightly elevated Fermi level $E_F + eV_b$, $eV_b \ll k_B T$, compared to the rest of the

3

device, so that each forward propagating miniband state is populated with a probability proportional to the the density of states. The position of each injected electron is evolved using the semiclassical equation of motion Eq. 1 with specular boundary conditions at the edges. To mimic the non-local resistance $(V_M - V_R)/I_L$, we consider an ensemble of $N_L$ injected states and calculate $(N_M - N_R)/N_L$, where $N_M$ is the number of electrons entering contract $M$ (of width $w = 250$ nm), and $N_R$ is a smooth background, calculated as $N_R = \sum_{i=1}^{N_L} w/d_i$, where $d_i$ is the distance between consecutive skips along the boundary for the $i^{\text{th}}$ trajectory.

The relation between gate voltage and the Fermi level is $V_g = E_F/e + en(E_F)/C_g$ where $n(E_F)$ is the electron density and $C_g/e = 4\,(\text{moiré unit cell area})^{-1}\,(2.75\text{ V})^{-1}$ is the measured geometric capacitance.

## 2 Supplementary Text

### 2.1 Determination of the miniband structure

We calculate the superlattice minibands of a fully aligned (misalignment angle $\theta = 0$) graphene/h-BN heterostructure by numerically diagonalizing the Hamiltonian (*20, 40*),

$$\hat{H} = v\vec{p}\cdot\vec{\sigma} + (U_0^+ f_+ + U_0^- f_-) + \xi\sigma_3\left(U_3^+ f_- + U_3^- f_+\right) + \frac{\xi\vec{\sigma}}{b}\cdot\left[\vec{l}_z \times \nabla\left(U_1^+ f_- + u_1^- f_+\right)\right]. \quad \text{(S.2)}$$

Here $\sigma_i$ are Pauli matrices, acting on Bloch states $(\phi_{AK}, \phi_{BK})^T$ in the $K$ valley ($\xi = 1$) and $(\phi_{BK'}, -\phi_{AK'})^T$ in the $K'$ valley ($\xi = -1$), and $\hbar = 1$. Functions $f_\pm = \sum_m (\pm 1)^{m+\frac{1}{2}} e^{i\vec{b}_m\cdot\vec{r}}$ are written using the six shortest Bragg vectors of the moiré superlattice, $\vec{b}_{m=0,\cdots,5} = R_{m\pi/3}(0, b)$, where $R_\varphi$ describes anticlockwise rotation by angle $\varphi$, and we use $b = 0.053\,\text{Å}^{-1}$ for the fully aligned heterostructure. The first term in $\hat{H}$ is the Dirac Hamiltonian of unperturbed graphene, while the remaining terms describe the superlattice perturbation. Among these, the first term describes a simple potential modulation; the second a A-B sublattice asymmetry, locally imposed by the substrate; and the third a modulation of A-B hopping. The strength of

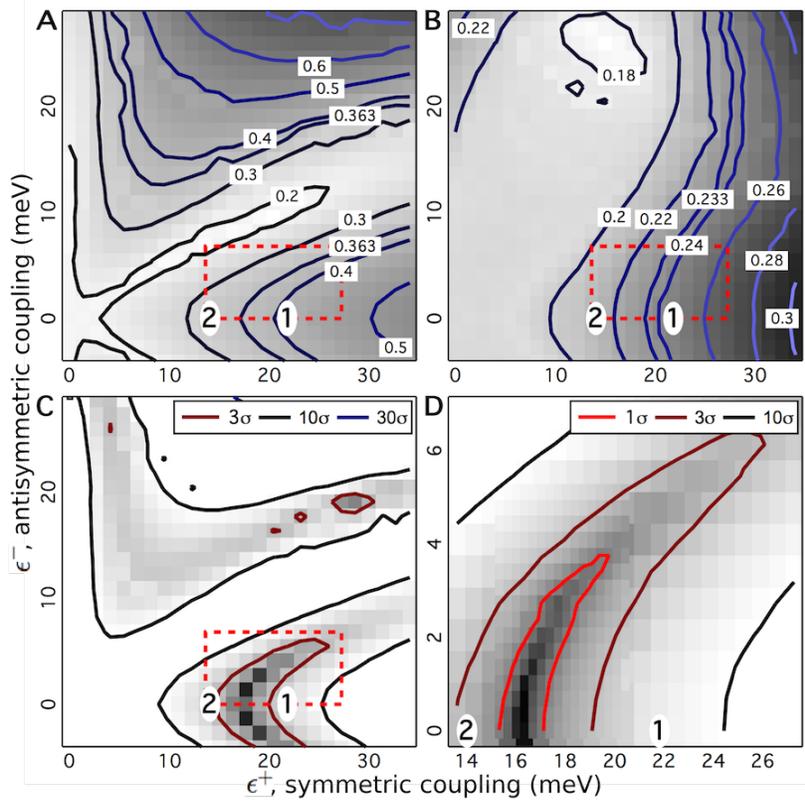

**Fig. S1.** (**A**) The prediction for the gate voltage ratio $\frac{V_v^1-V_v^2}{V_c^1-V_v^1}$ as a function of moiré perturbation parameters $\epsilon^+$ and $\epsilon^-$. The value observed in experiment is 0.363. "1" marks parameters used in (*20*), and "2" denotes parameters in (*41*). (**B**) Same as (A) for $\frac{V_c^2-V_c^1}{V_c^1-V_v^1}$. The observed value is 0.233. (**C**) Confidence regions for the estimation of parameters $(\epsilon^+, \epsilon^-)$, obtained by matching the calculated values in (A) and (B) with the experimentally observed values. Dark color is more likely as the prediction is more similar to experiment; contours are boundaries of confidence regions. (**D**) Zoom into the red dashed box in (A)-(C). The best choice of parameters is $(\epsilon^+, \epsilon^-) = (17\,\text{meV}, 0\,\text{meV})$.

each of these terms is characterized using parameters $U_{i=0,1,3}^{\pm}$, where $+/-$ is used for the part of each term which is symmetric/antisymmetric under the in-plane spatial inversion symmetry.

In principle, each parameter $U_i^{\pm}$ takes an arbitrary value. However, two microscopic models, one based on the hopping between the graphene and h-BN lattices (*19*), and the other on

scattering of graphene electrons off quadrupole electric moments in the hBN layer (*20*), predict

$$\{U_{i=0,1,3}^{\pm}\} = \epsilon^{\pm}\left\{\frac{\pm 1}{2}, -1, \frac{-\sqrt{3}}{2}\right\}. \tag{S.3}$$

By using Eq. (S.3) we reduce the number of parameters used to describe the superlattice perturbation from six in Eq. (S.2) to two ($\epsilon^{\pm}$) in Eq. (S.1). We treat $\epsilon^{\pm}$ as variable parameters, and establish their value with a fitting procedure using the experimentally measured TEF oscillation. First we note that the gate voltages $\{V_v^1, V_c^1, V_v^2, V_c^2\}$ at which TEF oscillations terminate (highlighted in Fig. 2B), correspond to saddle points in the miniband structure. Then, we compute the miniband structure resulting from each choice of ($\epsilon^+, \epsilon^-$), and compare its prediction for the gate voltage ratios $\frac{V_v^1-V_v^2}{V_c^1-V_v^1}$ and $\frac{V_c^2-V_c^1}{V_c^1-V_v^1}$ against the experimentally observed values. Figures S1A and B show the predicted gate voltage ratios $\frac{V_v^1-V_v^2}{V_c^1-V_v^1}$ and $\frac{V_c^2-V_c^1}{V_c^1-V_v^1}$ as a function of $\epsilon^+$ and $\epsilon^-$, which were calculated taking into proper account the geometric gate and quantum capacitances. We note that it is sufficient to consider only positive values of $\epsilon^{\pm}$ as the miniband structure is invariant under (i) $\epsilon^- \to -\epsilon^-$ and (ii) $(\epsilon^+, \epsilon^-) \to R(2\pi/3)(\epsilon^+, \epsilon^-)$ (*40*). Quantitative comparison to values measured in experiment constrains $\epsilon^{\pm}$ as shown in the form of confidence regions depicted in Figs. S1C and D. We visually estimate a standard error of $\Delta V_g = 0.03\,\text{V}$ in determining the gate voltage positions of a saddle point. The $1\sigma$ confidence region lies within $\epsilon^+ \approx 0.05 \cdot vb = 17\,\text{meV}$ and $|\epsilon^-| < 0.01 \cdot vb = 3\,\text{meV}$, and the match is best for $\epsilon^- \approx 0\,\text{meV}$. Hence we choose $(\epsilon^+, \epsilon^-) = (17\,\text{meV}, 0\,\text{meV})$ to calculate the miniband structure, electron dynamics, and TEF spectra shown in all figures in the main text.

Previous optical studies of graphene/h-BN heterostructures (*15, 41*) also employed model (S.3) for the superlattice perturbation, but assumed $\epsilon^- = 0$ *a priori*. Here, we have provided experimental justification for asserting $\epsilon^- = 0$ within this model, and provided a fit of $\epsilon^+ = 17\,\text{meV}$, which is comparable to previously estimated values (*15, 41*).

6

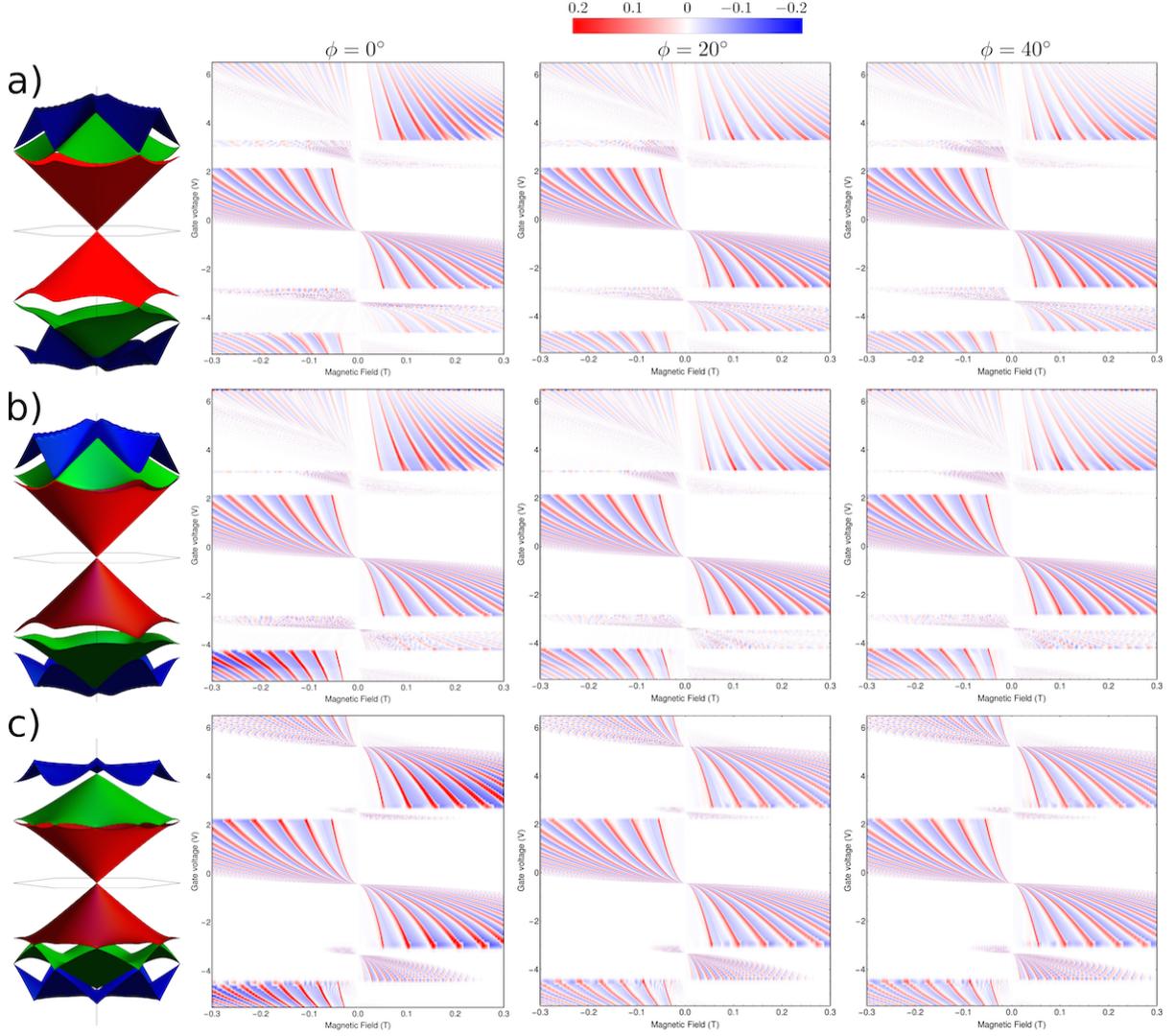

**Fig. S2.** The miniband structures and corresponding TEF spectra calculated for $\phi = 0°, 20°$, and $40°$. The superlattice parameters are chosen as either (**A**) $\epsilon^+ = 17\,\text{meV}$, $\epsilon^- = 0$, in accordance with the main text, (**B**) $U^+_{i=0,1,3} = \{-8.4, -10.7, -8.3\}\,\text{meV}$, $U^-_{i=0,1,3} = \{-5.6, 3.9, 3.4\}\,\text{meV}$ in accordance with Ref. (*42*) (a translation of the superlattice (*40*) was employed to emphasize the inversion symmetric part of the perturbation), or (**C**) $U^+_{i=0,1,3} = \{-52, 0, 0\}\,\text{meV}$, $U^-_{i=0,1,3} = \{0, 0, 0\}$, representing a scalar potential perturbation.

## 2.2 More examples of calculated TEF spectra

The left panels of Fig. S2 display further examples of miniband structures, calculated using Hamiltonian (S.2) and several different choices of the superlattice parameters $U^\pm_i$. The remain-

7

ing panels display the corresponding TEF spectra calculated for several choices of the angle $\phi$ (measured between moiré Bragg vector $\vec{b}_0$ and the direction perpendicular to the device edge). Importantly, the comparison of panels for various $\phi$ shows that the main features of the TEF spectra are independent of angle $\phi$ (the spectra will repeat after $\phi = 60°$). In particular, the gate voltages $\{V_v^1, V_c^1, V_v^2, V_c^2\}$ at which the TEF oscillations in a given miniband terminate, are set by the energy of saddle points in the minibands, and do not depend on $\phi$.

## 2.3 Influence of electron-electron scattering on the temperature dependence of the visibility of the TEF oscillations

To model the temperature dependent decay of the TEF oscillations, we calculate the spread of a bunch of non-equilibrium electrons as they propagate from the collector to the emitter using a Boltzmann transport equation. We take an initial electron distribution, injected at time $t = 0$, with wavevectors concentrated in a small range of angles, to mimic the focused electrons near the caustic trajectory. After this, the role of the magnetic field is non-essential to our model, as we shall consider the spread of the electron distribution in the direction transverse to the cyclotron path (below described by coordinate $y$), while the overall propagation of its center of mass displaces along the segment of a cyclotron semicircle to $x = vt$. Also, we neglect the moiré perturbation, which formally limits this calculation to Fermi energies within about half the band width of the first miniband (densities corresponding to C1 and V1 in the measurements).

Then, the Boltzmann transport equation reads (*44, 45*),

$$[\partial_t + v\sin(\theta_1)\partial_y] f(\vec{k}_1) = I\{f(\vec{k}_1)\}, \tag{S.4}$$

$$I\{f(\vec{k}_1)\} = \frac{8\pi|W|^2}{(2\pi)^4 v} \int d\vec{k}_2 d\vec{k}_3 \delta(k_1 + k_2 - k_3 - k_4) g_{1,3} g_{2,4} F,$$

$$F = \left[ -f(\vec{k}_1) f(\vec{k}_2)(1 - f(\vec{k}_3))(1 - f(\vec{k}_4)) + f(\vec{k}_3) f(\vec{k}_4)(1 - f(\vec{k}_1))(1 - f(\vec{k}_2)) \right].$$

Here $\theta_i = \arctan(k_i^y/k_i^x)$, $\vec{k}_i = (k_i^x, k_i^y)$, the distribution function $f(\vec{k}_1)$ depends implicitly on

8

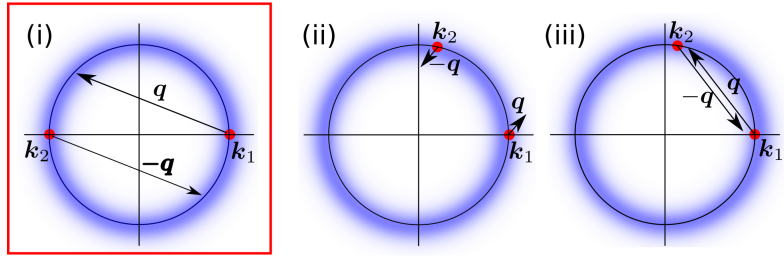

**Fig. S3.** The three possible ways for initial electron states (red) with wavevectors $\vec{k}_1$ and $\vec{k}_2$ ($\vec{q} = \vec{k}_3 - \vec{k}_1$) to scatter while conserving energy and momentum, and for which all initial and final states lie close to the Fermi line (blue). Only the process (i) can scatter the initial electron states through arbitrary angles.

y, $\vec{k}_4 = \vec{k}_1 + \vec{k}_2 - \vec{k}_3$ is determined by momentum conservation, $k_i = |\vec{k}_i|$, $g_{i,j=1,\cdots 4} = \frac{1+\cos(\theta_i - \theta_j)}{2}$ are the chirality factors (43), $W = \frac{\pi v}{2k_F}$ (where $k_F$ is the Fermi wavevector) is equivalent to the contact electron-electron potential $V(\vec{r}_1 - \vec{r}_2) = W\delta(\vec{r}_1 - \vec{r}_2)$, and we have taken into account spin-valley degeneracy.

To evaluate $I\{f(\vec{k}_1)\}$ in Eq. (S.4) we approximate,

$$f(\vec{k}_i) \approx \frac{1}{\exp[\frac{v\Delta k_i}{k_b T}] + 1} + \frac{\delta\mu(\theta_i)}{2k_B T}\frac{1}{1 + \cosh(\frac{\Delta k_i}{k_B T})} \tag{S.5}$$

where $\Delta k_i = k_F - k_i$, the Boltzmann constant is $k_B$, and $\delta\mu(\theta_i) \ll k_B T$ is a small, angle dependent, shift in the chemical potential attributed to electrons with momenta orientated along $\theta_i$.

For this distribution $I\{f(\vec{k}_1)\}$ is sharply peaked when $\Delta k_i/k_F \lesssim k_B T/(v k_F) \ll 1$ for each $i = 1, \cdots 4$. Also, by momentum conservation,

$$\Delta k_4 \approx k_F \left( \sqrt{3 + 2\cos(\theta_2 - \theta_1) - 2\cos(\theta_3 - \theta_1) - 2\cos(\theta_3 - \theta_2)} - 1 \right),$$

so that possible choices of $\theta_2$ and $\theta_3$ which satisfy $\Delta k_4/k_F \ll 1$ are divided into the three cases displayed in Fig. S3: either (i) $\theta_2 = \theta_1 + \pi + \Delta\theta_2$, or (ii) $\theta_3 = \theta_1 + \Delta\theta_3$, or (iii) $\theta_2 = \theta_3 + \Delta\theta_2$, where $\Delta\theta_i \lesssim k_B T/v k_F$. For process (i) the initial pair of electrons can scatter into

any state near the Fermi line, where as for cases (ii) and (iii) the angle of the scattered electron only deviates from that of an initial state by an amount $\lesssim k_B T / v k_F$. Because of this the contributions of processes (ii) and (iii) towards the relaxation rates for electron distributions are $\sim (m k_B T / v k_F)^2$ times lower than that generated by process (i) (46), which will be studied below ($m$ is the index of the angular harmonic).

By concentrating on process (i), our focused non-equilibrium electron distribution will typically decay by producing a beam of holes propagating in the opposite direction. As the holes separate quickly from the electron bunch, their effect on the decay of the electron distribution can be neglected. Hence we use $\delta\mu(\theta_2) = \delta\mu(\theta_4) = 0$ which reduces Eq. (S.4) to,

$$[\partial_t + v\sin(\theta_1)\partial_y]\,\delta\mu(\theta_1) = \frac{|W|^2}{32 k_B T \pi^3} \int dk_1 d\vec{k}_2 d\vec{k}_3 \delta(k_3 - k_1 + k_4 - k_2) g_{1,3} g_{2,4} \frac{\delta\mu(\theta_3) - \delta\mu(\theta_1)}{\prod_{i=1,\cdots 4} \cosh(\frac{v\Delta k_i}{2 k_B T})}.$$

Next, we expand $\delta\mu(\theta)$ in terms of its angular harmonics, $f(m)$,

$$\delta\mu(\theta) = \sum_m e^{im\theta} f(m),$$

so that the transport equation becomes,

$$\partial_t f(m) + \frac{v}{2i}\partial_y(f(m-1) - f(m+1)) = -\frac{f(m)}{\tau_m}. \tag{S.6}$$

Here we have introduced the relaxation time for the $m^{\text{th}}$ harmonic,

$$\frac{1}{\tau_m} = \frac{-|W|^2}{32\pi^3 k_B T} \int dk_1 dk_2 dk_3 d\tilde{\theta}_2 d\tilde{\theta}_3 \; k_2 k_3 \, \delta(k_1 + k_2 - k_3 - k_4) g_{1,3} g_{2,4} \frac{e^{im\tilde{\theta}_3} - 1}{\prod_{i=1,\cdots 4} \cosh(\frac{v\Delta k_i}{2 k_B T})} \tag{S.7}$$

and defined $\tilde{\theta}_3 = \theta_3 - \theta_1$, and $\tilde{\theta}_2 = \theta_2 - \theta_1$.

For process (i), $\tilde{\theta}_2 = \pi + \Delta\theta_2$ and the energy conserving delta function can be expressed as,

$$\delta(k_1 + k_2 - k_3 - k_4) = \frac{1}{k_F|\sin(\tilde{\theta}_3)|}\delta(\frac{\Delta k_1 + \Delta k_2 - 2\Delta k_3 - (\Delta k_2 - \Delta k_1)\cos(\tilde{\theta}_3)}{k_F \sin(\tilde{\theta}_3)} - \Delta\theta_2),$$

10

which simplifies the integration over $\tilde{\theta}_2$ and leads to,

$$\frac{1}{\tau_m} = \frac{(k_B T)^2}{v k_F} I(m), \tag{S.8}$$

$$I(m) \equiv \frac{1}{512\pi} \int d\xi_1 d\xi_2 d\xi_3 \, d\tilde{\theta}_3 \frac{\left[1 + \cos(\tilde{\theta}_3)\right]^2 \left[\cos(m\tilde{\theta}_3) - 1\right]}{|\sin(\tilde{\theta}_3)| \cosh(\frac{\xi_1 + \xi_2 - \xi_3}{2}) \prod_{i=1,\cdots 3} \cosh(\frac{\xi_i}{2})}$$
$$\approx \begin{cases} \alpha \log(\beta |m| + \gamma), & m \neq 0 \\ 0, & m = 0 \end{cases}$$

where $\alpha \approx 0.518$, $\beta \approx 2.28$, and $\gamma \approx -0.675$.

The collision integral increases with $m$, so that the spreading of a narrow beam of electrons is dominated by the relaxation of harmonics with $m \gg 1$. Hence, we treat $m$ as a continuous variable (and correspondingly allow $\theta$ to range between $\pm\infty$), and approximate $f(m-1) - f(m+1) \approx -2\partial_m$ in Eq. (S.6) to arrive at,

$$\partial_t f(t, y, m) - \frac{v}{i} \partial_y \partial_m f(t, y, m) = -\frac{1}{\tau_m} f(t, y, m),$$

where we explicitly list the transverse coordinate and time dependences in the distribution function. By taking the Fourier transform in the $y$-direction, $\hat{f}(t, q_y, m) = \frac{1}{2\pi} \int dy f(t, y, m) e^{-i q_y y}$, we obtain,

$$\partial_t \hat{f}(t, q_y, m) - v q_y \partial_m \hat{f}(t, q_y, m) = -\frac{1}{\tau_m} \hat{f}(t, q_y, m),$$

which is solved using

$$\hat{f}(t, q_y, m) = \hat{f}_0(q_y, m + v q_y t) e^{-\frac{G(m + v q_y t) - G(m)}{v q_y}}, \qquad \frac{dG(m)}{dm} = \frac{1}{\tau_m},$$

where $\hat{f}_0(q_y, m) = \hat{f}(t = 0, q_y, m)$. For simplicity, we assume the beam to be initially Gaussian in $y$, with characteristic width $a_e$ (the width of the emitter) and fully collimated, so that $\hat{f}_0(q_y, m) = (2\pi)^{-2} e^{-\frac{q_y^2 a_e^2}{2}}$, and,

$$f(t, y, \theta) = \frac{1}{(2\pi)^2} \int dq_y dm \, e^{-\frac{q_y^2 a_e^2}{2}} e^{-\frac{G(m + v q_y t) - G(m)}{v q_y}} e^{i q_y y} e^{i m \theta},$$

11

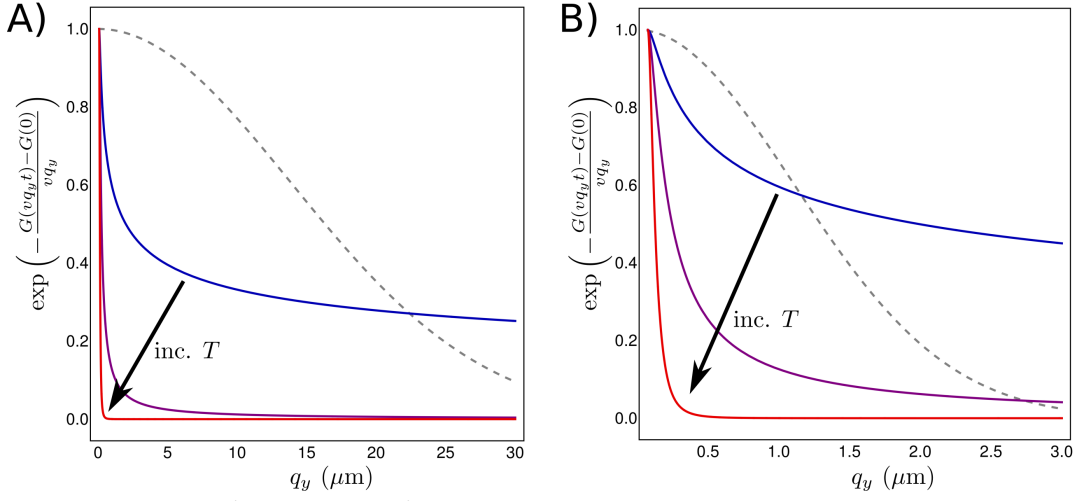

**Fig. S4.** Plots of $\exp\left(-\frac{G(vq_yt)-G(0)}{vq_y}\right)$ for $T = 0.5T_*$ (blue), $T = T_*$ (purple), and $T = 2T_*$ (red). Relaxation rates are taken from Eq. (S.8), $t = \pi L/(2v)$ and $L = 6.3\,\mu\text{m}$. The grey dashed line displays $\exp\left(\frac{-k^2A^2}{2}\right)$ for **(A)** $A = w/\sqrt{12}$ where $w = 0.25\,\mu\text{m}$ and **(B)** $A = L/(2\sqrt{12})$.

models the result of the transverse spreading of the beam after the time $t = \frac{\pi L}{2v}$ needed for it to reached the collecting electrode. To model the temperature dependent signal at the collector, $R(T)$, we take the convolution of the total distribution of electrons, $\int f(\frac{\pi L}{2v}, y, \theta)d\theta$, with a Gaussian weigh factor, $\exp(-y^2/2a_c^2)$, mimicking the collector width $a_c$, so that,

$$\frac{R(T)}{R(0)} = \frac{1}{\sqrt{2\pi}A}\int dq_y\, e^{-\frac{q_y^2 A^2}{2}} e^{-\frac{G(q_y\pi L/2)-G(0)}{vq_y}}, \tag{S.9}$$

$$= \frac{\sqrt{2}}{\sqrt{\pi}A}\int_{\frac{2(1-\gamma)}{\pi\beta L}}^{\infty} dq_y e^{-\frac{q_y^2 A^2}{2}} e^{-\frac{\alpha(k_B T)^2}{v^2 k_F}\left[(\frac{\pi L}{2}+\frac{\gamma}{\beta q_y})\log(\frac{\beta q_y \pi L}{2}+\gamma)-\frac{\pi L}{2}+\frac{1-\gamma}{\beta q_y}\right]},$$

$$A^2 = a_c^2 + a_e^2,$$

where lower cut off of $\frac{2(1-\gamma)}{\pi\beta L}$ for the integral is introduced to exclude unphysical harmonics with $m < 1$.

In Fig. S4 we display the factors in the integrand of Eq. (S.9) for various temperatures. When $T \lesssim T_*$, where

$$T_* = \sqrt{\frac{2v^2 k_F}{\alpha\pi L k_B^2}},$$

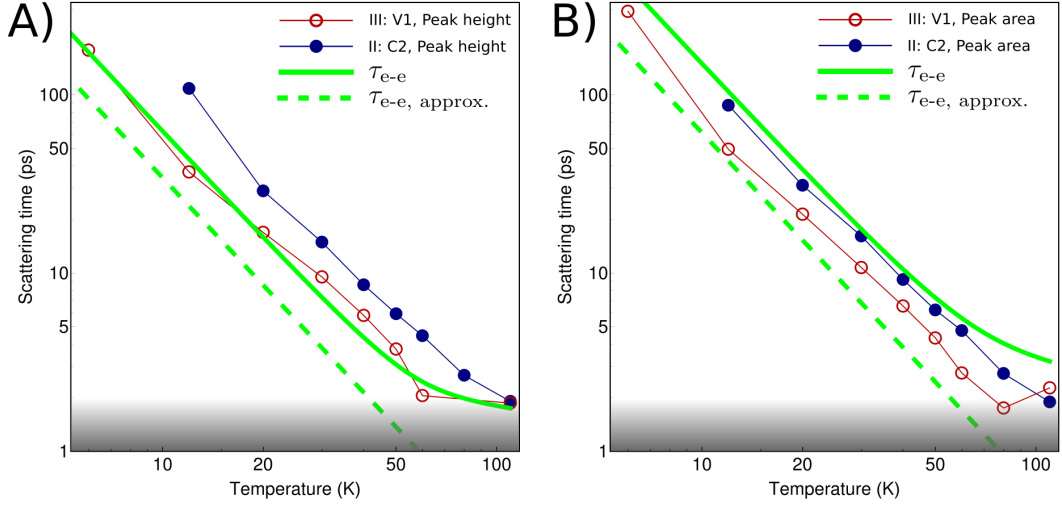

**Fig. S5.** (**A**) The effective scattering times, either extracted from the heights of the measured TEF peaks for the densities in bands C1 and V1 (red and blue circles) used in Fig. 4 of the main text, or, calculated using either the numerically evaluated form of $\tau_{\text{e-e}}$ (solid line) or its asymptotic Eq. (S.10) (dashed line) using $A = w/\sqrt{12}$ where $w = 0.25\,\mu$m. Note that the Fermi wavevector is almost identical for these two densities. (**B**) The same as (A) except extracting the scattering times from the measured TEF peak areas, and calculating them using $A = L/(2\sqrt{12})$. Shading is used to show the area for which the measured signal TEF has decayed too much for the scattering time to be reliably extracted.

the integral is dominated by the wide tails of the integrand, extending up to $q_y \lesssim A^{-1}$, where it decays exponentially with $L$ and $T^2$ (here $T_* \approx 45$ K). Hence for $T \lesssim T_*$, the decay of the TEF signal can be described using a scattering time, $\tau_{\text{e-e}}$, which we define similarly to that used for the experimental data in the main text,

$$\tau_{\text{e-e}}^{-1} \equiv -\frac{2v}{\pi L}\log\left(\frac{R(T)}{R(0)}\right) \approx \frac{\alpha(k_B T)^2}{v k_F}\log\left(\frac{\beta \pi L}{2A}\right). \quad \text{(S.10)}$$

Note that, this scattering rate is slower than the quasiparticle life-time calculated in Ref. (*43*).

For temperatures $T > T_*$, the tail of the integrand in Eq. (S.9) for large $q_y$ becomes suppressed, and the whole integral for $R(T)$ converges to a quantity whose dependence on $L$ and hence $t$ follows a power law rather than an exponential as a function of time. This crossover of

the signal decay with time leads to a slower decrease of $R(T)/R(0)$ with increasing temperature above $T_*$.

In Fig. S5 the numerically evaluated form of $\tau_\text{e-e}$ (solid green line) is compared to the approximate asymptotic behavior set by Eq. S.10 (dashed green line), and the scattering times extracted from either the measured TEF peak heights or TEF peak areas using the recipe described in the main text. For the calculated scattering time, we use $A = \frac{w}{\sqrt{12}}$ to model the decay of the peak heights (Fig. S5A), where $w = 0.25\,\mu$m is the width of the collector, and the $\sqrt{12}$ factor comes from matching the variance of the Gaussian distribution to that of a rectangular distribution. For Fig. S5B we take $A = \frac{L/2}{\sqrt{12}}$ because integrating the TEF signal to obtain the area of the peak, whose width is approximately half of the TEF periodicity, is then in turn approximately equivalent to adding up signal from a detector of width as large as $L/2$. In both cases the calculated and measured lifetimes are broadly comparable. Note that the calculated scattering time is only weakly (logarithmically) sensitive to the value of $A$ and therefore to whether we used $w$ or $L$ to estimate it.



\end{document}